# A minimum-hypothesis explanation for the "Radcliffe Wave": KH instability

Robert Fleck[1]

[1]Department of Physical Sciences, Embry-Riddle Aeronautical University, Daytona Beach, FL, USA

**Arising from: João Alves et al. *Nature* https://doi.org/10.1038/s41586-019-1874-z (2020)**

It has long been recognized that the gaseous and stellar components of the Milky Way and several other disk galaxies exhibit undulatory displacements ("corrugations") ranging from 1 to several kiloparsecs in wavelength and up to 350 parsecs in amplitude, and a variety of mechanisms, including gravitational instabilities, tidal interactions, collisions of high-velocity clouds, interaction of spiral density waves with the gaseous disk, and the undular mode of the Parker instability, have been proposed to account for these features[1,2]. Here it is suggested that the wavelike character of the recently discovered[3] 2.7-kiloparsec spatially and kinematically coherent complex of interstellar clouds in the solar neighbourhood—the "Radcliffe Wave"—may be the result of a Kelvin-Helmholtz instability (KHI) arising at the interface between the Galactic disk and non-corotating halo. Much like wind blowing over water, periodic waves will develop at the interface of fluids in relative motion[4,5]. Indeed, the fluid astronomical environment is ripe for KHI which has been shown to account for many and varied morphological and dynamical effects ranging in scale from comet tails to galaxies. The inferred oscillation about the Galactic midplane of the undulatory Radcliffe Wave is indicative of a wave-generating origin.

The maximum unstable wavelength for two superposed fluids rotating at angular velocity $\Omega$ is $\Lambda_{max} = \pi V_{rel}/\Omega$, where $V_{rel}$ is the relative streaming velocity of the two fluids[6]. The large-scale Galactic magnetic field $B \approx 2~\mu G$ aligned with this shearing flow in the solar neighborhood[7] will suppress the KHI if $V_{rel} \leq V_A = B(4\pi\rho)^{-1/2}$, the Alfvén speed in the region[5], where $\rho = \mu m_H n \approx 2.3 \times 10^{-25}$ g cm$^{-3}$ is the mass density of the gas having a particle density $n \approx 10^{-1}$ cm$^{-3}$ at the disk-halo interface one scale height $H$ ($\approx 200$ pc) above the Galactic midplane ($\mu = 1.38$ is the gas mean molecular weight and $m_H$ is the mass of the hydrogen atom). Thus the region will be KH unstable only if $V_{rel} \geq 20$ km s$^{-1}$. Measurements[8] of neutral hydrogen 21-cm emission reveal an *average* vertical falloff in the Galactic rotation curve approaching 30 km s$^{-1}$ kpc$^{-1}$ within 100 pc of the Galactic midplane, giving a velocity shear of slightly more than 10 km s$^{-1}$ across the disk ($2H$) for an instability driven from both sides of the midplane in this three-layer system. Accordingly, regions of the Galaxy experiencing this relatively low amount of shear should be stable against the KHI. However, this vertical velocity gradient varies across the Galaxy, rising to as much as 100 km s$^{-1}$ kpc$^{-1}$ in localized regions[8], and we can therefore expect KH-unstable regions wherever $V_{rel} \geq V_A \approx 20$ km s$^{-1}$. For $\Omega \approx 10^{-15}$ s$^{-1}$ in the solar neighbourhood, a shear of 20 km s$^{-1}$ gives $\Lambda_{max} \approx 2$ kpc, in agreement with the observed undulation of the Radcliffe Wave.

Simulations[9] of the effect of rotation on the KHI in protoplanetary disks show that the Coriolis force removes the stabilizing effect of velocity (and density) gradients on the KHI[5], and that the dominant unstable modes have wavelengths of several scale heights: $kH \sim 1$, where $k = 2\pi/\Lambda$ is the wavenumber of an unstable mode having wavelength $\Lambda$. Extrapolating this result to our Galaxy gives $\Lambda \approx 2\pi H \approx 1300$ pc, within the wavelength range established by $\Lambda_{max}$. The longest rise time for the instability $\approx (k_{min} V_{rel})^{-1} = \Lambda_{max}/2\pi V_{rel} = \frac{1}{2}\Omega \approx 20$ Myr, comfortably less than the roughly 200 Myr the Sun takes to orbit the Galaxy.

The KHI growth rate $\approx kV_{rel}$, so the shortest-wavelength unstable modes grow fastest. But viscosity effects[10] will stabilize wavelengths less than $2\pi (v^2/\varepsilon g)^{1/3}$, where[11] $v \approx d^2\Omega \approx 4 \times 10^{26}$ cm$^2$ s$^{-1}$ is the turbulent viscosity in the unstable layer above the midplane having a thickness $d \approx H$, $g = 3.5 \times 10^{-9}$ cm s$^{-2}$ is the disk surface gravity at one scale height above the midplane, and $\varepsilon \equiv (n_d - n_h)/(n_d + n_h) \approx 1$ for a disk-to-halo gas particle density ratio $n_d/n_h \gg 1$. (It can be shown[12] that kinematic viscosity and diffusion effects are several orders of magnitude less than this turbulent viscosity for the problem at hand.) Thus, viscosity will stabilize wavelengths less than about 700 pc, narrowing the range of unstable wavelengths to $\approx 1 - 2$ kpc, matching very well the undulation scale of the Radcliffe Wave. In any case, no unstable modes will have wavelengths less than the full disk thickness ($2H \approx 400$ pc).

Physically, KHI occurs when the aerodynamic suction $P \approx \rho_h V_{rel}^2$ above (and below) the disk exceeds the disk gravitational restoring force per unit area.[13] Setting the vertical pressure gradient $\sim P/A \approx \rho_h V_{rel}^2/A$ across an unstable wave of amplitude $A$ equal to the volume gravitational restoring force $\rho_d g$ gives $V_{rel} \approx 40$ km s$^{-1}$ for $\rho_h/\rho_d \approx 1/10$ and a Radcliffe Wave amplitude $A \approx 160$pc, well within the range of plausible shearing velocities. The amplitude of the instability will be limited by the onset of turbulence accompanying the development of the nonlinear KHI (Kelvin-Helmholtz "rollover"), and, in any event, by the vertical scale height $H \approx 200$ pc.

In the spirit of William of Ockham (c. 1285 – 1347), a KHI explanation for the undulatory Radcliffe Wave hinges on a *single* (hence "minimum-hypothesis") determining parameter, $V_{rel}$, however uncertain its actual value may be. It would be of interest to measure velocity gradients in the region of the Radcliffe Wave in order to refine the proposed KHI model developed here, and also to extend the model's reach to other similar features within our and other galaxies.